# On turbulence: deciphering a renormalization flow out of an elliptic curve[1]

LUÍS G.C.D. BORGES [2]

The use of power series to model phenomena in nature, no matter what its specific details look, is an old and common practice. Sometimes, however, when the practical situation at hand imposes us to acknowledge our own limits, in dealing with the infinite, then, so to speak, one is driven to approximations, and that is when the convenience of using a polynomial model may make it appear as a natural, somewhat alternative way to go ahead.

That was also the moment when, having become interested in studying the dynamics of those formal power series associated with a few, special elliptic curves over $\mathbb{Q}$, most notably, those ones with semi-stable reduction, I turned my attention to some certain ninth degree polynomials deducible from them. It was around 1997. The choice of these curves, just by the way, was motivated by their being both modular and isomorphic to rigid analytic varieties, a circumstance that seemed, and still seems to me to make them quite interesting, notwithstanding the deep technicalities involved, way ahead my own mathematical competence, so far. The reason for this interest will be made somewhat clearer in the following lines.

Anyway, taking these steps, I was following ideas suggested by my readings of John T. Tate's 1974 paper "The Arithmetic of Elliptic Curves" (in Inventiones math. 23, 179-206) and also Joseph H. Silverman's 1986 book, with the same title (from Springer's GTM collection, n.106). My goal was to develop models for turbulence phenomena; more precisely, to use those very polynomials to model the renormalization flow with scale as its parameter that is typical of that kind of phenomena. I say, coherent structures being replicated across the scales, like little vortices becoming bigger and bigger, or smaller and smaller, and so on and so forth.

---

[2] Author's contact e-mail: luiscborges@netcabo.pt .

So, there I was, probing the dynamics of some polynomial objects (see fig. 1), quite uncertain about what the right choice of an observable would be, with which to put up a classification, discriminating them in a suitable way.

$$P(x) = x^3(1 + A_1x + A_2x^2 + \cdots + A_6x^6)$$

(with each $A_n$ ($n = 1, 2, \ldots, 6$) denoting a polynomial in the coefficients of the Weierstrass model for a semi-stable elliptic curve over $\mathbb{Q}$)

Fig. 1

During the first years of this research, it was around 2004 that the topological entropy, as inferred from the symbolic dynamics codifying the sequence of escape times along an orbit, started to look like a promising option. In fact, the computer simulations made in those years suggested that, in about 80% of the cases studied, the conductor values of the curves giving origin to the diverse polynomials are values that do acknowledge for a simple, yet non trivial ordinal relation across the set of families constituted by these later. This, of course, insofar as that relation concerns the values numerically found for their topological entropy, and up to a reasonable point: the conductor of a curve is an important invariant and what the simulations showed was that, as measured by the topological entropy, the polynomials deduced from curves with equal conductor (within the same isogeny class, for instance) tend to behave consistently, against a gauge defined by a curve with another conductor value.

Proceeding with this percentage in mind, after some 6 years of successive failures in bringing about what would be an obviously needed, more conclusive result, the topological entropy still seems a good choice for an observable, among others. Meanwhile, however, another line of investigation was to be developed, which is to explore the idea that at least some of the dynamical properties of these polynomials might, or better, may indeed be encoded in the L functions that come, so to speak, with the former, original elliptic curves. Enough said about the modular form bias in my line of reasoning, here, it boils down to the claim, certainly yet to be proved sound, that these L functions somehow encode, or more strongly, do determine the dynamics of the polynomials with which I proposed myself to model turbulence.

This was, and is, I say, a long shot indeed, moreover, to be fired by the most non experienced hands! Yet, as the experts have pointed out, that deep arithmetic information is encoded in the behavior of these L functions beyond their region of convergence (a notice I took from Anthony W. Knapp's 1992 book, "Elliptic Curves", from Princeton Univ. Press, Mathematical Notes, vol. 40), I was aiming at a closer, preliminary target, assuming the hypothesis that the same goes, and is to be learned from the region of divergence, in what concerns the relevant dynamics.

Thus, I decided myself to look at the way that the many iterates generated from a sample of 30 L functions explode to infinity, in this later region. More precisely, from a sample which was chosen in a mixed, systematic random way, among the population of all semi-stable elliptic curves over $\mathbb{Q}$, with conductors up to 1000. In other words, taken from Robert L. Devaney's 2006 work on Complex Exponential Dynamics (published 2010 in the Handbook of Dynamical Systems, vol. 3, edited by H. Broer et al.), I turned my attention to some few bifurcation diagrams, with my curiosity being partially aroused by the beautiful images thereby exposed (see fig. 2, for my own, pale images of the original, R. L. Devaney's ones).

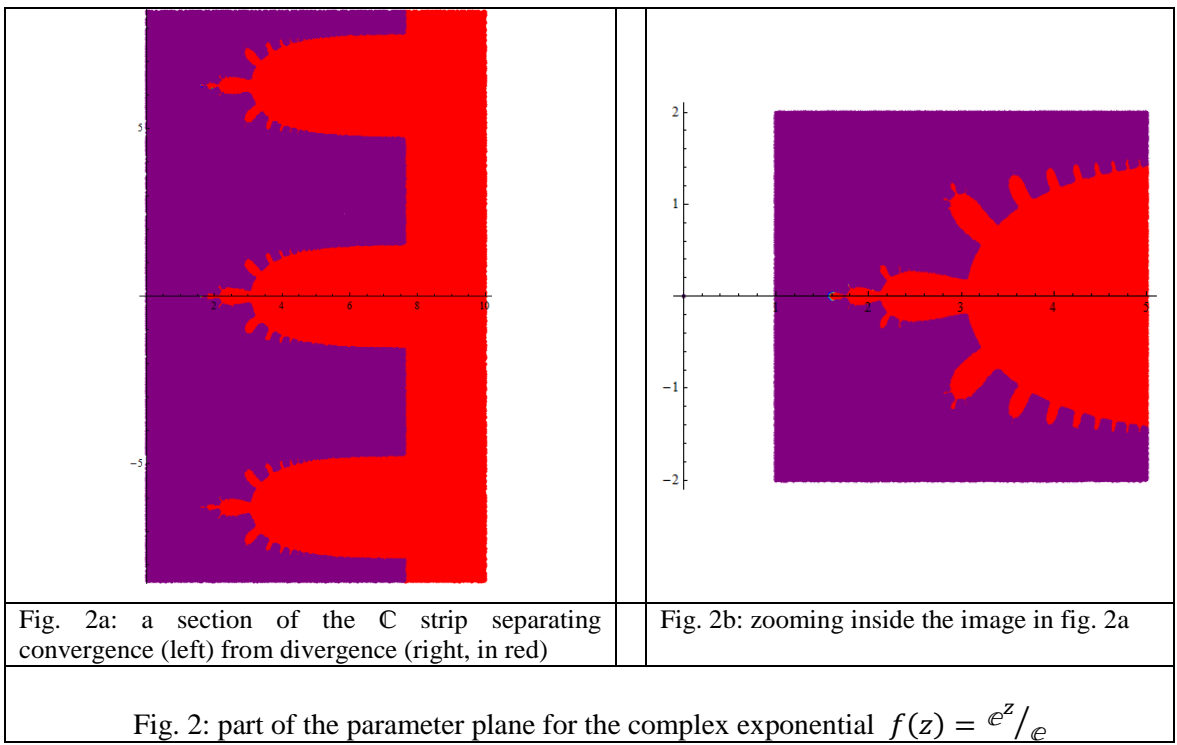

| | |
|---|---|
| Fig. 2a: a section of the $\mathbb{C}$ strip separating convergence (left) from divergence (right, in red) | Fig. 2b: zooming inside the image in fig. 2a |

Fig. 2: part of the parameter plane for the complex exponential $f(z) = {e^z}/{e}$

The plan was to apply the same technique, namely, an escape times algorithm, to render images detailed enough to probe for the existence of some analogous structuring, in the bifurcation diagram for an L function.

Not an easy task. However, relying heavily, and confidently, upon the algorithms made available by J. E. Cremona, in his 1997 2$^{nd}$ edition of "Algorithms for Modular Elliptic Curves", published by Cambridge Univ. Press, and using Mathematica$^{®}$, from Wolfram Research, I did manage to write the codes needed to generate the first thousand coefficients in the series representation of each of these, so involved L functions. Then, as these codes started running, by the end of 2008, my surprise gradually unfolded itself into astonishment, as the following sequence of images of bifurcation diagrams being unveiled started to be documented (see Figs. 3).

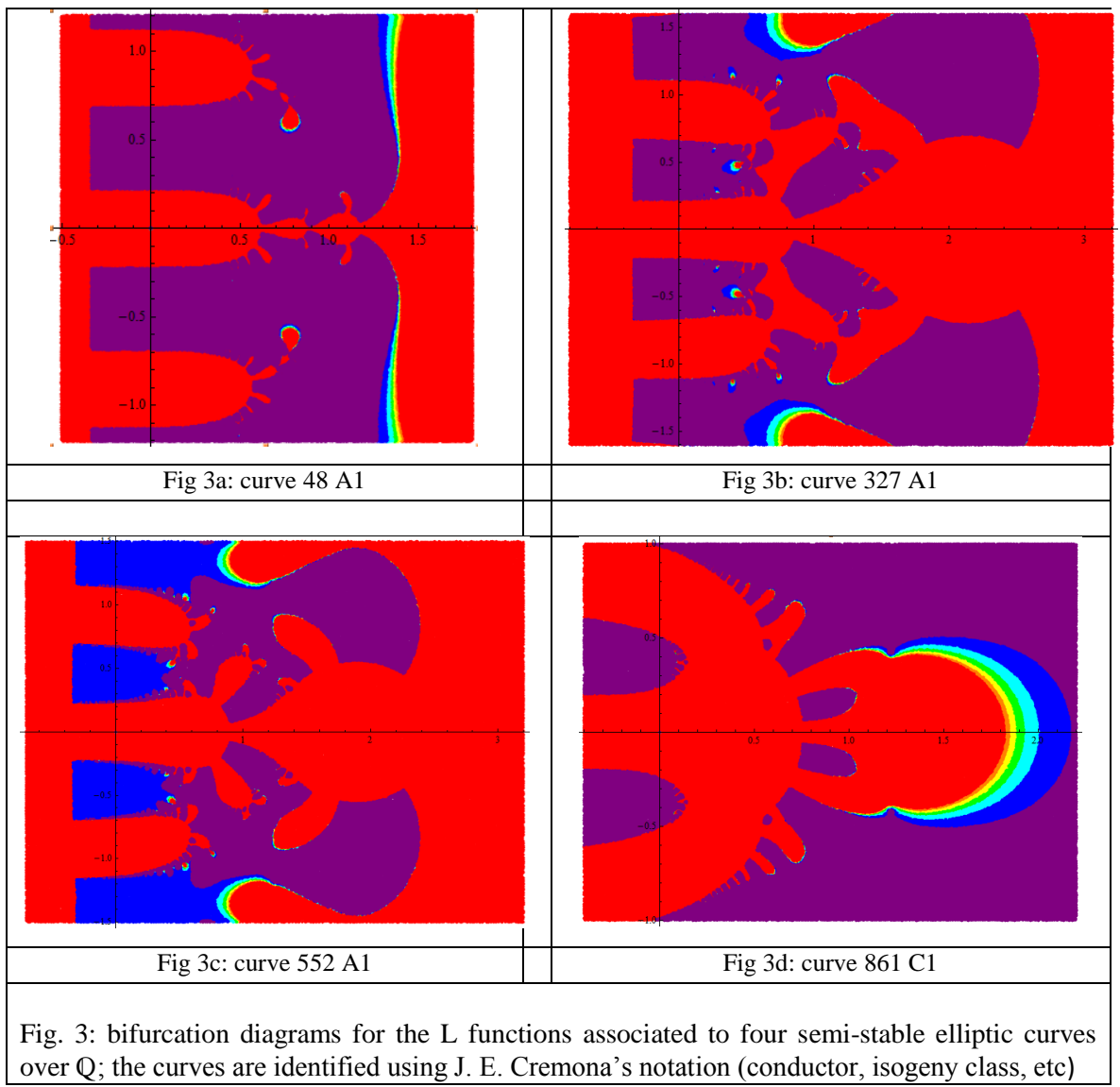

| Fig 3a: curve 48 A1 | Fig 3b: curve 327 A1 |
| Fig 3c: curve 552 A1 | Fig 3d: curve 861 C1 |

Fig. 3: bifurcation diagrams for the L functions associated to four semi-stable elliptic curves over ℚ; the curves are identified using J. E. Cremona's notation (conductor, isogeny class, etc)

Indeed, it was obvious that I had come across something worth of a more systematic exploration then, say, just the production, or the rendering of interesting pictures. Before going further, however, a few inferences were in order and may still justify immediate acceptance, as a more close look into what happens with curve 21A1, together with the previous images (in Fig.3), seems to follow by inspection only (see next, Fig. 4).

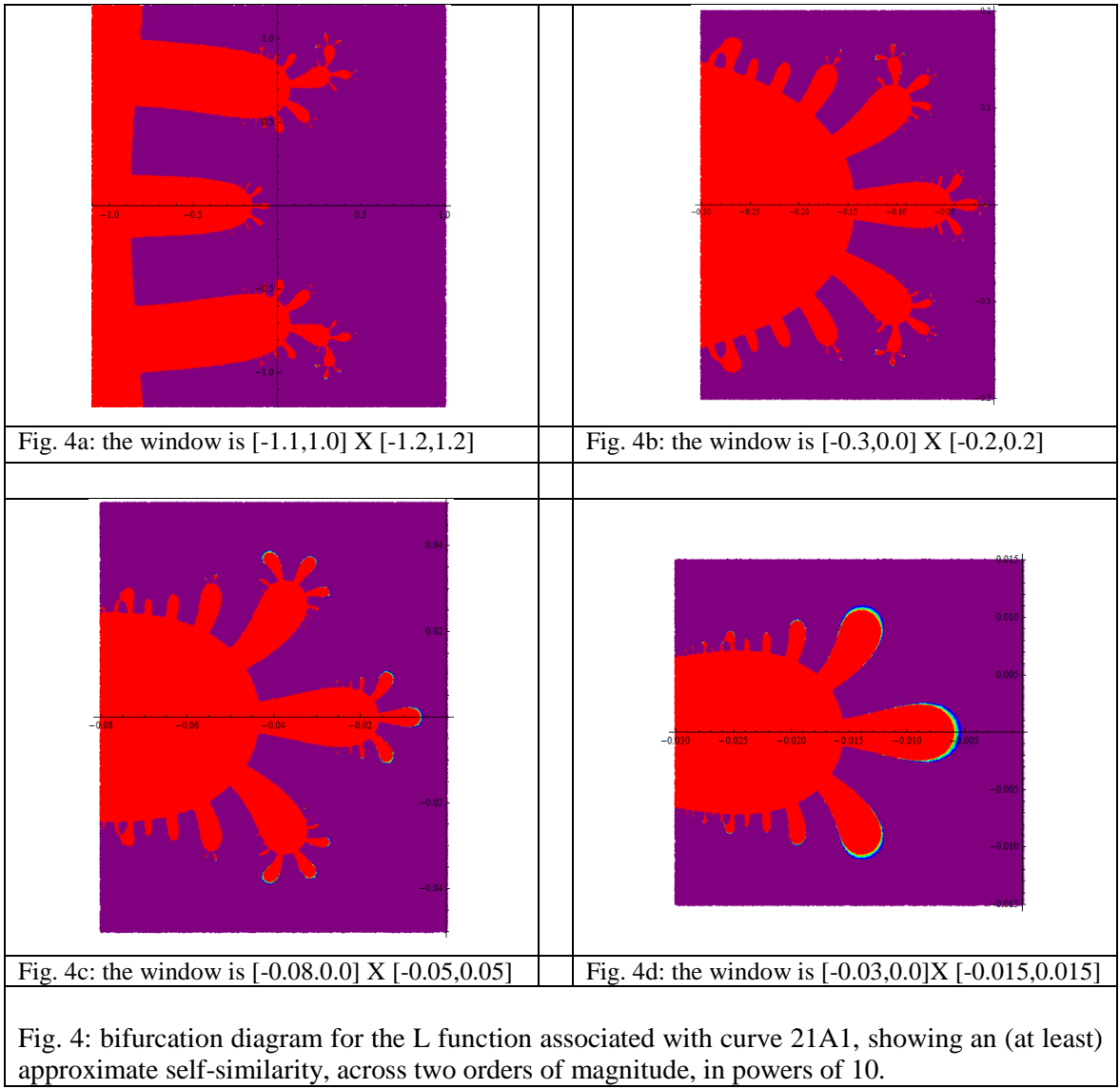

| Fig. 4a: the window is [-1.1,1.0] X [-1.2,1.2] | Fig. 4b: the window is [-0.3,0.0] X [-0.2,0.2] |
| --- | --- |
| Fig. 4c: the window is [-0.08.0.0] X [-0.05,0.05] | Fig. 4d: the window is [-0.03,0.0]X [-0.015,0.015] |

Fig. 4: bifurcation diagram for the L function associated with curve 21A1, showing an (at least) approximate self-similarity, across two orders of magnitude, in powers of 10.

In fact, taking R. L. Devaney's cited work as a reference, it is now possible to resume the following inferences:

a) We no longer necessarily have a vertical strip separating a region of convergence from one of divergence, as it is with the complex exponential; we can have divergence on both sides of the strip (figs 3a, 3b and 3c);

b) On the other hand, similarly to what happens in the complex exponential case, each extreme of at least some of these protuberances shown (the 'fingers', in R. L. Devaney's terms), making the interface between convergence and divergence, may have in its own boundary an image of the whole strip ( the boundary of the central, main protuberance shown in fig. 4b reproduces the image shown in fig. 4a, apart from the main curvature);

c) The particular aspects that are unveiled in the approximate self-similarity shown, from curve to curve, or L function to L function, depend on the coefficients taken into consideration in the numerical approximations that were made to these later functions: the more coefficients one takes into account, the more structure will appear in the bifurcation diagrams, beyond some correspondent or convenient scale threshold.

It may be noticed, with appreciation, that this last inference can be translated into the following, more sophisticated assertion: the never ending flow of structure, or renormalization flow that one will find, scanning these bifurcation diagrams across the scales, is something totally encoded (or better still encrypted) in the algorithms which model the action of some certain Hecke operators on the Hilbert space spanned by the modular newforms that, according to the modular curve 'conjecture', share these L functions with the curves before-mentioned. This is because the Fourier coefficients of these later functions are to, or can be deduced from the eigenvalues of those former operators.

Anyway, the question was, it still is, how to deduce, from these L functions, whatever may help in the understanding of the behavior of some associated ninth degree polynomials. Now, given that this behavior is assumed to model a few dynamical aspects of turbulence, close in essence to what one may call renormalization flows, it follows as quite a natural option to start with the dynamical behavior of these very functions. I say, going beyond the image rendering of their bifurcation diagrams and making its dynamics the object of some convenient quantification.

Thus, starting by the end of 2008 and up to the end of 2010, a long trial and error process was to be developed, that took me to try with the escape rate, $\tau$, as an observable. The escape rate is a statistical quantification of how fast the trajectories of a map, in this case, the iterates of an L function explode to infinity. An idea that was inspired in the 1987 paper by T. Tél, "Escape rate from strange sets as an eigenvalue" (published in Phys. Rev. A, 36, 3, 1502-1505) and is further developed in the 1993 book by C. Beck and F. Schlögl, "Thermodynamics of Chaotic Systems", edited by Cambridge Univ. Press.

More to the point, the escape rate, $\tau$, is related to a special value of the topological pressure which, in turn, is a quantity to be interpreted as a free energy for the map under consideration. Thus, the vague conjecture started to take form, that a variation in free energy, along the above mentioned sample of 30 L functions, might be related, in a systematic way, to a correspondent variation in topological entropy within the set of the associated ninth degree polynomials.

Given the harsh experience of the many previous failures, however, I decided not to adventure further, at least before granting myself this approach to be consistent with some canon, within elliptic curve theory. Then, making what I suppose could honestly be called a very wild guess (notwithstanding what the Birch and Swinnerton-Dyer conjecture tells us), I ran the codes previously written for the L functions, in order to make numerical approximations to the values taken by them on the point $z = 1$, in the complex plane.

By the end of December 2010, the two lists of values thereby obtained along the same sample (one for $L(1)$, the other for $\tau$) were studied in search of a connection: the ordinal correlation coefficient (Spearman coefficient, $r_S$) was determined and found to be $r_S = -0.76$ . Finally, this result was tested for statistical significance. For a sample of dimension 30, with 28 degrees of freedom, on a level of confidence $\alpha = 0.001$ , the null hypothesis that there is no correlation between the two lists was rejected (the critical point for this test being $T_{cr} = 0.45$).

In conclusion, it is reasonable to argue for the case that these notices, hereby given, do invite for the planning of a bigger scale investigation on the questions that were raised, or to which they relate. Should a common life-time span be enough to complete it or not, the fact is that the comprehension of turbulence still lies way ahead, in many a researcher's dream, just as it also drives one to discoveries unexpected, surely worth the effort… and real. What a great challenge, then, for one to stand up to and strive, developing mathematics!

These were my thoughts, back in February 2011. A lot of work was done since then, but it was my conviction that they were worth publishing, just as they were, almost a year ago. Unfortunately, the manuscript was not given the justice of an acknowledgment of reception, from the editor to whom it was sent, not even after an e-mail was sent, in August 2011, giving notice of my deception. Thus, on past December 9, a letter was handwritten withdrawing it from submission.

Meanwhile, as work went on, the choice was made of another observable with reference to which one can measure the polynomial dynamics above mentioned. As some preliminary results looked promising, I took another sample which was chosen as before, in a mixed, systematic random way, among the population of all semi-stable elliptic curves over $\mathbb{Q}$, with conductors now up to 200 000 (taken from J.E. Cremona's kept web page – last consulted in August-, at the University of Warwick, [http://www.warwick.ac.uk/~masgaj/ftp/data/index.html](http://www.warwick.ac.uk/~masgaj/ftp/data/index.html) ). Just as with the first sample, the same, somewhat unexpected connection was found, in the form of a statistically significant ordinal correlation, reaching a level of confidence $\alpha = 0.001,$ between the escape rates of their L-functions and the associated values on the point the point $z = 1$, in the complex plane.

An extensive set of computations started then, last Summer, which already sums up to some thousand hours running the programs which simulate these dynamics and their measuring processes. This will take time, but I do hope to be able to announce some results soon.

Lisbon, 1/1/2012